\documentstyle[preprint,pre,aps,epsfig]{revtex}
\begin{document}
\draft
\title{Magnetization Reversal in Ferromagnetic Film Through Solitons 
by Electromagnetic Field}
\author{V. Veerakumar\,\,and\,\, M. Daniel$^{\thanks {nld@bdu.ernet.in}}$}
\address{The Abdus Salam International Centre for Theoretical Physics, Trieste, Italy}
\address{\mbox{and}}
\address{Centre for Nonlinear Dynamics, Department of Physics, Bharathidasan 
University, Tiruchirappalli 620 024, India} 
\maketitle 
\begin{abstract}
We study the reversal of magnetization in an isotropic ferromagnetic 
film free from charges by exposing it to a circularly 
polarized electromagnetic (EM) field. The magnetization excitations 
obtained in the 
form of line and lump solitons of 
the completely integrable modified KP-II equation which is derived 
using a reductive perturbation method from the set of coupled 
Landau-Lifshitz and Maxwell equations. It is observed that when the polarization 
of the EM-field is reversed followed by a rotation, for every 
$ {\pi \over 2} $-degrees, the magnetization is reversed.
\end{abstract}
\pacs{PACS: 75.60.jk, 05.45.Yv, 02.30.Ik, 41.20.-q,}
\maketitle

Studies on developing high density magneto-optic data storage and development
of ultrafast magnetic recording have attained momentum due to their high technological 
implications in the very recent times.  It has been identified both experimentally 
and theoretically that magnetization reversal in magnetic films is one 
of the very fundamental issues in magnetic data storage 
\cite{bac00,acr00,bau00a,bau00b}. 
In conventional magnetic recording the reversing field is applied anti-parallel to the 
direction of magnetization of the medium in which case the reversal takes place at the 
nano-second level. However, much shorter reversal times can be achieved through 
precessional reversal of the magnetization \cite{bac00}.  The torque developed 
between the magnetic moment of the medium and the external magnetic field will 
make the magnetic moment to precess at the pico-second scale.  The precessional 
motion is governed by the Landau-Lifshitz (LL) equation and recent experiments 
have shown the validity of the equation at the pico-second level \cite{bac00}. 
Thus the switching process or magnetization reversal can be understood by 
solving the Landau-Lifshitz equation of motion for the applied magnetic field.  
In this context static and time dependent (pulse) applied field switching 
phenomenon have been studied in the past \cite{bau00a,bau00b,bro63}.  
Magnetization reversal 
studies in the case of a more general applied magnetic field that varies both 
spatially and temporally is also equally interesting and it needs attention.  In the 
context of magneto-optics, it is very relevant and important to consider the 
applied field as the magnetic field component of the EM-field and thus the 
problem can be formulated in terms of the LL equation coupled with the 
Maxwell equations. 
The one-dimensional version of this kind of problems have been studied recently in 
the case of isotropic and anisotropic ferromagnets and soliton modes were found 
to represent the magnetization excitations and 
the EM-field has also been modulated in the form of solitons 
\cite{veera1,veera2,veera3,kraenk}.
The purpose of the present paper is to examine the reversal process of magnetization in a 
ferromagnetic film by generating soliton modes during precession of 
magnetization in the presence of an EM-field. By assuming that there are no free 
electric charges in the medium, we solve analytically the LL equation 
coupled with Maxwell equations in two spatial dimensions using a reductive perturbation method.  

The dynamics of magnetization density  
in an isotropic charge-free ferromagnetic 
film under the influence of an external EM-field 
can be expressed in terms of the LL equation 
\cite{landau}
$ {\partial}_t {\bf M}$ $ = {\bf M} \wedge \left [ \nabla^2
{\bf M} + A{\bf H} \right ]$,~~$ {\bf M}^2 =1$, 
where $ {\bf M}(x,y,t) = (M^x, M^y, M^z) $ is the magnetization density
and ${\bf H}(x,y,t) = (H^x, H^y, H^z) $ is the magnetic field component 
of the electromagnetic field. 
The first term in the right hand side of the LL equation represents 
the contribution due to spin-spin
exchange interaction between the nearest neighbours
and the term proportional to $ A $ ($= g\mu_B$; g = gyromagnetic ratio, 
$\mu_B$ = Bohr magneton) corresponds to 
the interaction between the magnetization of the medium and the magnetic 
field component of the EM-field 
(Zeeman term).
When $ {\bf H} $ is a constant or a time dependent field 
set along a specific direction, it can be transformed away using
the transformation $ {\bf M}^{\pm}$ $\equiv (M^x \pm iM^y)$ 
$= {\bf{\tilde M}}^{\pm} e^{\mp i \int_{-\infty}^t {\bf H}(t') dt'}$ 
and the dynamics remains 
the same as in the case without any external field. However, when the 
field ${\bf H} $ varies spatially, the effect due to the field cannot  
be transformed away in this fashion.  It may be noted that in the LL equation we have not 
included the phenomenological Gilbert damping term 
because we have assumed that the medium does 
not contain any free charges which on interaction with magnetic electrons
will introduce damping. 

The interaction between the magnetic field component of the EM-field and matter or 
material medium can be expressed in terms of Maxwell equations 
\cite{jack}. In the absence of static 
and moving charges, Maxwell equations can be written as
$ \nabla^2 {\bf H}$ $- {\bf \nabla}
({\bf \nabla}.{\bf H})$ $ = {1 \over c^2}$ $ {\partial}_t^2 
\left[ {\bf H} + {\bf M} \right] $, 
where $ {\bf \nabla}(= {\hat{x}} \partial_x + {\hat{y}} \partial_y )$ 
and $\nabla^2 (= \partial_x^2 + \partial_y^2) $ are the two-dimensional gradient and 
Laplacian operators respectively, and $c={ 1 \over {\sqrt{\mu_0 \epsilon_0}}} $ is the velocity 
of propagation of the EMW and $ \epsilon_0 $ and $ \mu_0 $ are the dielectric 
constant and permeability of the medium respectively.  An untreated ferromagnetic material has 
the constitutive relation $ {\bf H} $ $={ {\bf B} \over \mu_0} - {\bf M} $ 
where $ {\bf B}(x,y,t) $ $=(B^x,B^y,B^z) $
is the magnetic induction. Now using this constitutive relation, 
the LL and Maxwell equations can be rewritten as 
\begin{mathletters}
\label{model}
\begin{eqnarray}
{\partial}_t {\bf M}  &=& {\bf M} \wedge \left [ \nabla^2
{\bf M} + {A \over \mu_0}{\bf B} \right ], \qquad {\bf M}^2 =1 , \label{eq3} \\
{\tilde{\Delta}} {\bf B} 
&=& {1\over \epsilon_0}\left[ {\bf \nabla} 
({\bf \nabla}\cdot {\bf M}) - \nabla^2 {\bf M} \right ]. \label{eq1}
\end{eqnarray}
\end{mathletters}
where the operator $ {\tilde{\Delta}} $ 
$= ({\partial}_t^2 - c^2\nabla^2 ) $. 
Thus, the set of coupled equations (\ref{eq3}) and (\ref{eq1}) completely
describe the interaction of the EM-field with the ferromagnetic film 
and the dynamics of the associated fields namely magnetization, 
magnetic induction and magnetic field from the EM-field.

The nonlinear character of the LL equation (\ref{eq3}) makes the 
analysis difficult in the present form and hence
we try to solve Eqs.(\ref{eq3}) and 
(\ref{eq1}) using a reductive perturbation method 
\cite{tanuit}. 
Assuming that the plane wave of the EM-field travels along $x$-direction in the 
ferromagnetic film ($xy$-plane) we introduce the wave 
variable $ {\hat{\xi}} = (x-vt) $ where $ v $ is the group velocity. 
We also introduce slow space and time variables through the 
stretching $ \xi = \varepsilon {\hat{\xi}} $, $ \zeta = \varepsilon^2 y $ and 
$ \tau = \varepsilon^3 t $, where $ \varepsilon $ is the small perturbation 
parameter.  The solutions of Eqs.(\ref{model}) are then expanded asymptotically 
about uniform values  
\begin{equation}
\label{eq4}
{\bf F}(\xi,\zeta,\tau) = {\bf F}_0 (\xi,\zeta,\tau) + 
\varepsilon {\bf F}_1(\xi,\zeta,\tau) + \varepsilon^2 {\bf F}_2 (\xi,\zeta,\tau)
+...,
\end{equation}
where {\bf F} stands for the magnetic induction ${\bf B} $ and 
the magnetization $ {\bf M} $. We substitute the slow variables 
and the expansions in the component forms 
of Eqs.(\ref{eq3}) and (\ref{eq1}), 
collect terms corresponding to similar 
powers of $ \varepsilon $ and solve the equations at different orders.  
At $ O(\varepsilon^0) $ from the component forms of Maxwell 
equations (Eq.(\ref{eq1})), we obtain the results
$ B^x_0 = 0 $, $ B^{\alpha}_0 = k M^{\alpha}_0 $,  where $\alpha =y,z$ and  
$ k = [\epsilon_0(c^2-v^2)]^{-1} $ and from the LL equation (Eq.(\ref{eq3})) 
after using the above results obtained we get $ M_0^x = 0 $. 
The results at $O(\varepsilon^0)$ thus show that a planar
anisotropy is developed in the magnetization $(M^y_0 - M^z_0)$, 
the magnetic induction $(B^y_0 - B^z_0)$ and in the magnetic field 
$(H^y_0 - H^z_0) $ space 
at the lowest order of 
expansion. 
At $ O(\varepsilon^1) $, from Eq.(\ref{eq1}) 
we obtain  $ B^x_1 = 0 $, $ B^{\alpha}_1= k M^{\alpha}_1 $ and from Eq.(\ref{eq3}) 
we find that 
$ {\partial}_{\xi} M^y_0 $ $= {kA \over v} M^z_0 M^x_1 $ and
$ {\partial}_{\xi} M^z_0 $ $= -{kA \over v} M^y_0 M^x_1 $.  
At  $ O (\varepsilon^2) $, after using the results from 
the lower orders, we find from Eq.(\ref{eq1}) that  
$ B^x_2 = 0 $  and
\begin{equation}
\label{eq6a}
\left [ B^{\alpha}_2 - k M^{\alpha}_2 \right ] 
= -{1 \over \epsilon_0 k} \left[ 2v 
{ \int_{-\infty}^{\xi} d\xi' 
{\partial}_{\tau} B^{\alpha}_0 }
+ c^2 { \int_{-\infty}^{\xi} d\xi' {  \int_{-\infty}^{\xi} d\xi'
{\partial}_{\zeta}^2 B^{\alpha}_0 } } \right],
\end{equation}
and from Eq.(\ref{eq3}), we get
\begin{equation}
\label{eq6b}
{\partial}_{\xi} M^x_1 = {1 \over v}
\left\{ M^z_0 {\partial}_{\xi}^2 M^y_0 - 
M^y_0 {\partial}_{\xi}^2 M^z_0 - A
\left[ M^y_0 B^z_2  + M^y_2 B^z_0 
- M^z_2 B^y_0  - M^z_0 B^y_2 \right]
\right \}.
\end{equation}
We now try to 
solve the set of equations (\ref{eq6a}) and (\ref{eq6b}) to evaluate the 
complete set of solutions at the lowest nonvanishing order.  For this 
it is advantageous to switch to the polar coordinate representation. 
As the results at $ O(\varepsilon^0) $ show that 
the magnetic field is restricted to the $(H^y_0 -H^z_0)$ plane, we 
consider a circularly polarized or rotating magnetic field component of the 
EM-field in 
this plane by considering $ {\bf H}_0 = (0, sin\theta, cos\theta) $. This
makes us also to choose
$ {\bf M}_0 $ $= (0,sin\theta , cos\theta) $, 
where $ \theta $ = $ \theta(\xi, \zeta, \tau) $
is the angle made between the direction of 
the applied EM-field and the uniform magnetization of the film.
Thus Eq.(\ref{eq6b}) in the above polar co-ordinate system 
after using Eq.(\ref{eq6a})  becomes
\begin{eqnarray}
\mu {\partial}_{\xi}^2 \theta &=&
3 \gamma
\left[ sin\theta \int_{-\infty}^{\xi} d\xi' \int_{-\infty}^{\xi} d\xi'
{\partial}_{\zeta}^2 cos\theta 
- cos\theta \int_{-\infty}^{\xi} d\xi' \int_{-\infty}^{\xi} d\xi'
{\partial}_{\zeta}^2 sin\theta \right]
\nonumber \\
&+&  {2vA \over \epsilon_0} \left[
sin\theta \int_{-\infty}^{\xi} d{\xi}'
{\partial}_{\tau} cos\theta  
- cos\theta \int_{-\infty}^{\xi} d\xi'
{\partial}_{\tau} sin\theta 
\right], \label{eq8}
\end{eqnarray}
where $ \mu = ( {v \over kA} - 1) $ and $ \gamma = {c^2 A \over 3 \epsilon_0} $. 
Differentiating Eq.(\ref{eq8}) twice with respect to $ \xi $, we get
\begin{eqnarray}
\label{eq10}
{\partial}_{\xi} 
\left[ { F(\theta) \over {\partial}_{\xi} \theta  } \right] {\partial}_{\xi} \theta
&=& -\mu ({\partial}_{\xi}^2 \theta )
{\partial}_{\xi} \theta +
3 \gamma \left[
cos\theta \int_{-\infty}^{\xi} d\xi'
{\partial}_{\zeta}^2 cos\theta 
+ sin\theta \int_{-\infty}^{\xi} d\xi'
{\partial}_{\zeta}^2 sin\theta 
\right]
\nonumber \\
&+& 3 \gamma {\partial}_{\xi}  
\left[ sin\theta \int_{-\infty}^{\xi} d\xi'
{\partial}_{\zeta}^2 cos\theta  
- cos\theta \int_{-\infty}^{\xi} d\xi'
{\partial}_{\zeta}^2 sin\theta \right], \label{eq9}
\end{eqnarray}
where 
$ F(\theta) $ $={\partial}_{\tau} \theta $ $+ \mu
{\partial}_{\xi}^3 \theta $.. While writing the above equation, we have rescaled 
$ \tau \rightarrow {\epsilon_0 \over 2vA} \tau $.
We assume that the fields vary very slowly along $y$-direction 
when compared to $x$-direction (i.e.), 
(${\partial}_{\zeta}^2 \theta \ll {\partial}_{\xi} \theta $). 
Then we substitute Eq.(\ref{eq8}) in the resultant equation 
obtained after differentiating Eq.(\ref{eq10}) once with 
respect to $ \xi $ and again after successive integration and 
differentiation we finally obtain  
$\partial_{\xi} 
\left[ {\partial}_{\tau} f \right. $ $- {3\over2} \mu f^2{\partial}_{\xi} f $
$ \left.+ \mu {\partial}_{\xi}^3 f \right] $ 
$= - 3\gamma {\partial}_{\zeta}^2 f $
$+ 3\gamma\left[ {\partial}_{\xi}^2 f 
+{\partial}_{\xi} f {\partial}_{\zeta} f \right. $
$+ \left.{\partial}_{\xi}^2f \int_{-\infty}^{\xi} d\xi' {\partial}_{\zeta} f  
\right] $, where  
$ f ={\partial}_{\xi} \theta $.  
When $ \gamma = \mu  $,  this is equivalent to the completely 
integrable modified Kadomtsev-Petviashvili (MKP) 
equation 
\begin{equation}
{\partial}_{\tau} f  + 
\mu {\partial}_{\xi}^3 f -
3 \mu \left[{1 \over 2} f^2 {\partial}_{\xi} f 
- {\partial}_{\zeta} w  + w { \partial}_{\xi} f  
\right]=0, \qquad
{\partial}_{\xi} w ={\partial}_{\zeta} f. 
\label{eq12} 
\end{equation}
Eq.(\ref{eq12}) when $\mu = i $ and  $\mu = 1 $ are respectively 
known as the MKP-I
and MKP-II equations.  However, in our problem $ \mu $ cannot be 
imaginary and hence we have
only the MKP-II equation for our further analysis.
Different types of soliton solutions to the MKP-II equation 
such as line soliton, lump soliton and breather 
have been found 
using the ${\bar{\partial}}$-dressing and inverse scattering transform (IST) 
methods \cite{kono2}. 
However, as the structure of breather solution is very rich we are 
unable to present them here. The general N-line soliton solution 
is obtained using 
${\bar{\partial}}$-dressing method \cite{kono2} and
the simplest line soliton of the MKP-II equation corresponding to $N=1 $ can 
be explicitly written in the form
$ f(\xi, \zeta, \tau)$ $= -2(\alpha_1 - \beta_1)^2 \gamma_1 / $ $ 
[\alpha_1 \beta_1^2 ({e^{-G} - {\alpha_1 \gamma_1 \over \beta_1 } e^G)(e^{-G} 
- \gamma_1 e^G)]}$,
where $ 2G = ({1 \over \alpha_1} - {1 \over \beta_1}) \xi $ $-
({1 \over \alpha_1^2} - {1 \over \beta_1^2}) \zeta $ $
-4({1 \over \alpha_1^3} - {1 \over \beta_1^3}) \tau $ 
$+ ln2|{R \over {\beta_1 - \alpha_1}}| $
and $ \gamma_1 = sgn({R \over {\beta_1 - \alpha_1}}) $.  Here $ R $ is the Kernel and 
$ \alpha_1 $ and $ \beta_1 $ are arbitrary real constants used in the IST analysis. 
This solution is nonsingular only
if $  \gamma_1 < 0 $ and $ \alpha_1, {1 \over \beta_1} > 0 $.
The line is regular in $ \xi $ and $ \zeta $ and is a constant along a particular 
direction.  

Lumps are rational solutions which normally decay in all directions in the plane. 
Unlike the MKP-I case, in the case of MKP-II, the rational solutions of the 
equation are singular.  For instance, the simplest ($N=1$) lump soliton of MKP-II equation 
corresponding to N=1 can be written as 
$ f(\xi,\zeta,\tau)$ $=  2 \alpha_1 /[{ {\alpha_1^2 \over 4} -  (\xi 
+ { 2\zeta \over \alpha_1} - {12 \tau \over \alpha_1^2} })^2 ]$, 
where  $ \alpha_1 $ is an arbitrary real constant.
This solution describes the uniform motion of two simple poles 
of opposite signs along a line  which are parallel to each other 
with a distance $ \alpha_1 $ and move with equal velocity
$ 12 \alpha_1^{-2} $. 

Using the above line and lump solitons 
in the relation
$ f = {\partial \theta \over \partial \xi} $, $ \theta $ 
and finally after using in the relation connecting $M^x_1, M^y_0 $
and $ M^z_0 $ we can calculate the components of magnetization at the 
lowest existing order.
For example, the line soliton of the x-component of magnetization
at the lowest existing order ($M^x_1$) is found to be \cite{kono2}
\begin{equation}
\label{line}
M^x_1 = {-2v(\alpha_1 - \beta_1)^2 \gamma_1 \over 
kA \left\{ \alpha_1 \beta_1^2 {e^{-G} - {\alpha_1 \gamma_1 \over \beta_1 } e^G e^{-G}
- \gamma_1 e^G} \right\} } ,
\end{equation}
Fig.(1a) shows a snapshot  
of $M^x_1$-line soliton at $ \tau = 1 $ unit for 
$ v = 0.5 $, $ k = 0.25 $, $ A = 1.0 $, $ \alpha_1= 5.0 $, 
$ \beta_1 = 2.5 $ and $ \gamma_1=-0.5 $.
\begin{figure}[h]
\begin{center}
\includegraphics[width=\columnwidth]{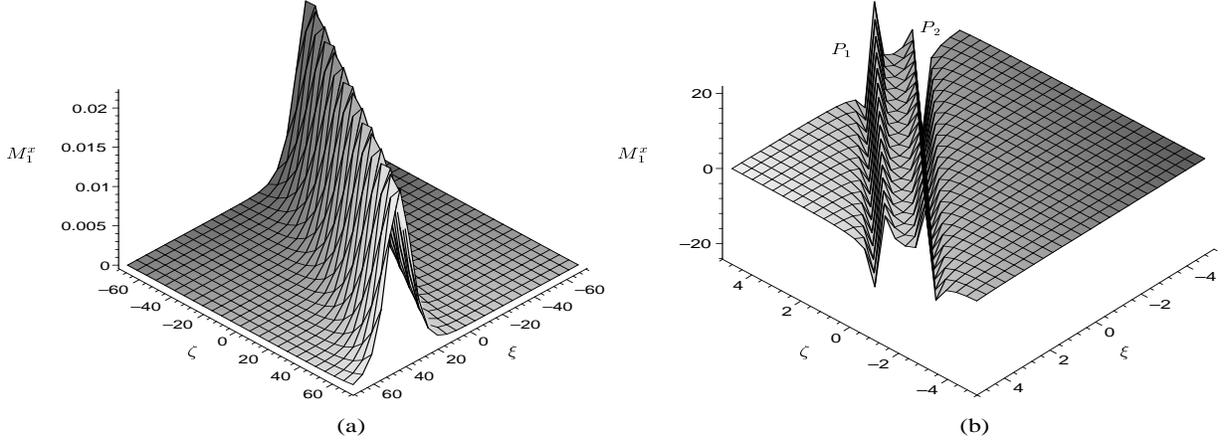}
\end{center}
\caption{
(a) Line soliton of Magnetization ($M^x_1$) at $ {\tau =1} $ unit for $ v= 0.5 $, 
$ k= 0.25 $, $ A = 1.0 $,  
$ \alpha_1 = 5.0 $, $\beta_1 = 2.5$ and $\gamma_1 = - 0.5$, 
(b) Lump soliton of magnetization ($M^x_1$) at 
$\tau = 1 $ unit for $ v = 0.5 $, $ k = 0.25 $, 
$ A = 1.0 $ and $ \alpha_1 = 2.05 $.}
\end{figure}

The lump 1-soliton solution of the x-component of magnetization ($M^x_1$) is obtained as
\begin{equation}
\label{lump}
M^x_1 = { 2 v\alpha_1 \over { kA \left\{ {\alpha_1^2 \over 4} -  (\xi 
+ { 2\zeta \over \alpha_1} - {12 \tau \over \alpha_1^2} )^2 \right\} } }.
\end{equation}
In Fig.(1b), a snap shot of the lump soliton of the  
$x$-component of the magnetization for 
$ v = 0.5 $, $ k= 0.25 $, $ A = 1.0 $ and $ \alpha_1 = 2.05 $ at $\tau = 1 $ is given.
In Fig.(1b) the uniform motion of the two simple poles, $ P_1 $ and $ P_2 $
separated by a distance of $ 2.05 $ units and travelling with a velocity of $ 2.855 $
units can be observed.
Similarly, line and lump soliton 
solitons for the $y$ and $z$ components of magnetization can also be constructed.

The magnetization states observed in the form of line and lump solitons in the previous 
case correspond to the states when the rotating circularly polarized 
magnetic field component of the EM-field is acting on the ferromagnetic film which is 
magnetized perpendicular to the film at the lowest order. It is interesting to find that 
when the direction of polarization of the magnetic field component of the 
external EM-field is reversed (i.e. if the initial field is left circularly polarized, 
now in the present case it should be right circularly polarized and vice versa) 
and then  rotated continuously,  
for every $ {n\pi \over 2 } $, $ n=0,1,2,...$ degrees of  
rotation, the direction of magnetization in the film is reversed.  
In other words, the above can be 
achieved by transforming $ v \rightarrow -v $ and 
$ \theta \rightarrow \theta + n {\pi \over 2} $. 
In view of these transformations, we now have 
$ {\bf M}_0$ $ = (0,cos\theta,sin\theta)$. 
We use this and the above transformations and 
also the results from Eq.(\ref{eq6a}) in Eq.(\ref{eq6b}). 
And after repeating the same calculations of 
the previous case as found after Eq.(\ref{eq6b}) we once again
obtain the MKP-II equation (\ref{eq12}). 
Then we calculate the line and lump solitons of the 
$x$-component of the magnetization $M^x_1$ using the
relation connecting
$ M^y_0 $, $ M^z_0 $  as done before
and the results are found to be 
$ M^x_1 $ $= 2v(\alpha_1 - \beta_1)^2 \gamma_1 $ $ / 
[ kA \{ \alpha_1 \beta_1^2 ({e^{-G} - {\alpha_1 \gamma_1 \over \beta_1 } e^G)(e^{-G}
- \gamma_1 e^G)}\} ] $ for the line soliton 
and the lump soliton in the form 
$ M^x_1 $ $ = -2 v\alpha_1 / $ ${ kA \{ {\alpha_1^2 \over 4} -  (\xi 
+ { 2\zeta \over \alpha_1} - {12 \tau \over \alpha_1^2} )^2 \} } ] $.
On comparing the line and lump solitons for $M^x_1$ 
in both the cases, it can be observed that there is a sign change in the 
expression for solitons and hence  
magnetization in the present case is reversed.  
\begin{figure}
\begin{center}
\includegraphics[width=\columnwidth]{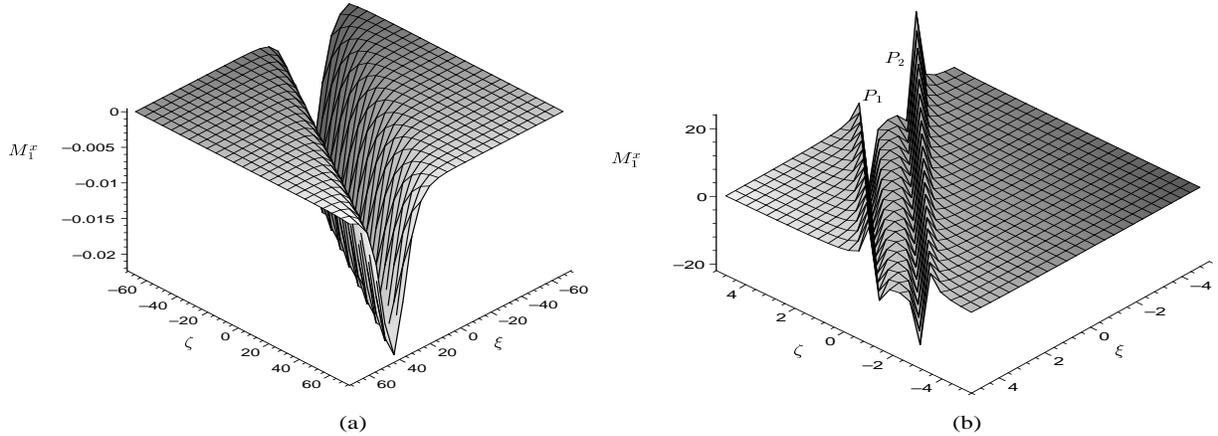}
\end{center}
\caption{
(a) Reversal of line soliton of the magnetization ($M^x_1$)
given in Fig.(1a) for the same values of the parameters, 
(b)  Reversal of lump soliton  of the mangnetization 
($M^x_1$)
given in Fig.(1b) for the same values of the parameters. Both are
due to change in direction 
of polarization (left $\leftrightarrow $ right) followed by 
the rotation of the magnetic field by ${\pi \over 2}$ degrees.}
\end{figure}
The new configurations of magnetization due to reversal in the 
case of $M^x_1 $ corresponding to the line soliton and also 
lump soliton with reference to the two poles $P_1$ and $P_2 $ 
have been demonstrated in Figs.(2a) and 
(2b) respectively. Thus the reversing of the direction of polarization of the magnetic 
field and a rotation by $ {\pi \over 2} $-degrees reverses the original 
magnetization states given in Figs.(1a,b). 
As the precessional frequency of the magnetic dipole moment in ferromagnets is 
in the pico-seconds scale, the magnetization reversal can take place in a faster rate
compared to the conventional reversal process that takes place in the nano-seconds scale 
by the reversal of magnetization of the domains.
This has also close correspondence with 
the recent experimental results found in \cite{bac00} on in-plane magnetized cobalt films in which 
pulse-like time dependent magnetic field in the plane of the film as short as 
two pico-seconds is able to reverse the magnetization.
It was found that when the 
field encompasses right angle to the magnetization of the film the reversal can 
be triggered by even very small fields. Also, our results have very close 
correspondence with the numerical analysis of the magnetization reversal 
found in refs.\cite{bau00a,bau00b} using 
four pico-seconds magnetic field pulse which has been explained based on the LL 
equation with 
Gilbert damping also to take into account the relaxation..

In this paper we have studied the reversal of magnetization in an isotropic charge-free 
ferromagnetic film when an EM-field is applied on it, by solving the coupled 
Maxwell equations and LL equation in two-spatial dimensions using a 
reductive perturbation 
method.  The results show that at the lowest order of perturbation the ferromagnetic 
film is magnetized normal to the plane of the film.  In the next order of perturbation 
it is found that the magnetization is excited and a coherent magnetization structure in 
the form of line and lump solitons of the MKP-II equation is obtained. 
Interestingly we found that, when the direction of 
polarization of the applied EM-field is reversed  
and the field is rotated continuously, the 
magnetization gets reversed for every addition of ${\pi \over 2}$-degrees.  
As this effects the magnetization reversal via the precessional motion, it will decrease
the reversal time to the order of pico-seconds than nano-seconds in conventional 
magnetization reversal processes.
This phenomenon has very close correspondence with the recent experimental and 
numerical observations of magnetization reversal in cobalt film when ultrashort 
magnetic pulses are applied to it.  This interesting phenomenon foresee applications
in ultrafast magnetic recording in future.

This work was done within the framework of the Associateship Scheme of the 
Abdus Salam International Centre for Theoretical Physics, Trieste, Italy. 
V.V acknowledges  CSIR for financial support in the form of a Senior Research
Fellowship. The work of M.D forms part of a major DST project.

\end{document}